\documentclass[ twocolumn,superscriptaddress,bibnotes,amsmath,amssymb,aps]{revtex4-2}
\usepackage{graphicx}
\usepackage{amssymb,amsmath}
\usepackage{bm}
\usepackage{dcolumn} 
\usepackage{subfigure}
\usepackage{float}
\usepackage[OT1]{fontenc} 
\usepackage{slashed}
\usepackage{color}
\usepackage{verbatim}
\usepackage{times}
\usepackage{graphicx}
\usepackage{url} 
\usepackage[colorlinks,linkcolor=red,anchorcolor=green,citecolor=blue,hypertexnames=true]{hyperref}
\usepackage{ dsfont }
\usepackage{natbib}
\usepackage{bm}

\begin{document}
\title{ Variational  quantum state diagonalization with computational-basis probabilities} 
\author{Juan Yao}
\email{juanyao.physics@gmail.com}
\affiliation{International Quantum Academy, Shenzhen 518048, Guangdong, China}
	
\begin{abstract}
In this report, we propose a novel quantum diagonalization algorithm based on the optimization of variational quantum circuits.  
Diagonalizing a quantum state is a fundamental yet computationally challenging task in quantum information science, especially as the system size increases.  
To address this challenge, we reformulate the problem as a variational optimization process, where parameterized quantum circuits are trained to transform the input state into a diagonal form.  
To guide the optimization, we develop two objective functions based on measurement outcomes in the computational basis.  
The first objective function utilizes global computational basis probabilities, with the optimized value directly yielding the purity of the input state.  
The second objective function, designed for enhanced experimental feasibility, is constructed solely from single-qubit probabilities. It admits an elegant and compact analytical form that significantly reduces the exponential measurement complexity, while still effectively driving the state toward a diagonal representation.
Through numerical simulations and analytical insights, we demonstrate that our variational optimization framework successfully produces the diagonal form of an input quantum state, offering a scalable and practical solution for quantum state diagonalization.
\end{abstract}
	
\maketitle

\section{Introduction}
Diagonalization is a fundamental mathematical tool with widespread applications in mathematics~\cite{bookLA}, physics~\cite{bookQM}, and machine learning~\cite{bookML}. Despite its broad utility, classical diagonalization techniques often become computationally prohibitive for large-scale systems~\cite{QR_book,poweriter_BRONSON2014237,inverse,Numerical_book}.
Quantum computing offers a promising alternative, with the potential to solve certain problems exponentially faster than classical methods~\cite{Nielsen_Chuang_2010,shor1,shor2}. This motivates the development of quantum state diagonalization algorithms that leverage quantum resources to overcome classical limitations~\cite{PhysRevA.97.012327}.
Recently, variational quantum algorithms (VQAs) have emerged as a promising framework for quantum state diagonalization, particularly suited to noisy intermediate-scale quantum (NISQ) devices~\cite{VQC,PhysRevX.6.021043}. In these approaches, a parameterized quantum circuit is trained within a hybrid quantum-classical optimization loop to transform an input quantum state into a diagonal form.
A number of recent studies have explored this direction~\cite{QPCA,subasi_entanglement_2019, PhysRevA.101.062310, VQdia,VQSE}, demonstrating the feasibility of variational diagonalization.
For example, Ref.~\cite{VQdia} constructs a diagonalizing unitary using inner-product test circuits, which require two copies of the input density matrix. This method involves ancillary qubits, controlled-swap operations, and multiple copies of the quantum state, thereby demanding complex quantum resources and posing significant experimental challenges.
Ref.~\cite{VQSE}, reduces these requirements by using only a single copy of the input state per iteration. It defines the objective function as the energy expectation of a non-degenerate Hamiltonian. However, this strategy requires careful design of the Hamiltonian to ensure both non-degeneracy and diagonalization in the computational basis, which limits its general applicability and introduces additional complexity.
Moreover, most of these methods focus on partial diagonalization, such as extracting dominant eigenvalues or principal components, rather than achieving a full diagonal representation of the quantum state.

In this report, we propose a variational quantum diagonalization algorithm that overcomes the limitations of existing approaches by introducing an experimentally friendly objective function. Since the variational training process requires repeated evaluations of the objective, its efficiency and ease of implementation are critical.
Our objective function relies solely on measurement probabilities in the computational basis, avoiding the need for ancillary circuits, swap tests, or specially constructed non-degenerate Hamiltonians. We introduce two objective functions tailored to different experimental scenarios:
(i) a global objective function based on full-basis probabilities, which can directly capture properties such as state purity, and
(ii) a local objective function constructed entirely from single-qubit probabilities, which dramatically reduces the measurement complexity.
This makes our approach both theoretically scalable and practically feasible for current quantum hardware. It offers a practical route toward full quantum state diagonalization using variational methods on NISQ devices~\cite{Preskill2018quantumcomputingin, McClean_2016}.

\section{Result}
\subsection{global probabilities}
\begin{figure}[t]
\begin{centering}
\includegraphics[width=0.45\textwidth]{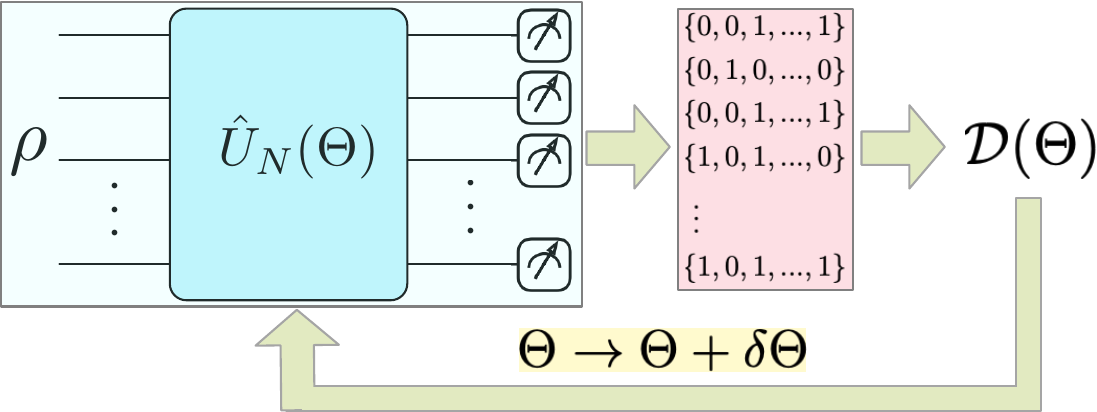}
\caption{ The diagonalization of the input quantum state is realized through optimizing a variational quantum circuit. An $N$-qubit quantum state $\rho$ is transformed using a parameterized unitary matrix $\hat{U}_{N}(\Theta)$.  The evolved state $\rho' = \hat{U}_{N} \rho \hat{U}_{N}^\dagger$ is followed by a computational-basis measurements, which contribute to the objective function. By optimizing the parameters $\Theta$ according to the proposed objective function, the evolved state $\rho'$ is automatically brought into a diagonal form.}
\label{Fig1}
\end{centering}
\end{figure}

As shown in the schematic diagram in Fig.~\ref{Fig1}, an $N$-qubit quantum state $\rho$ undergoes a parameterized unitary transformation $\hat{U}_N(\Theta)$, resulting in the evolved state $\rho' = \hat{U}_N(\Theta)\rho \hat{U}_N^\dagger(\Theta)$. At the end of the circuit, a sufficient number of measurements are performed on $\rho'$ in computational basis. Statistical analysis of the measurement outcomes provides the occupation probabilities for each  computational basis  with 
\begin{equation}
p_{q_1,q_2,...,q_N}  = {\rm Tr} \{\rho^\prime |q_1,q_2,... ,q_N\rangle \langle q_1,q_2, ...,q_N|\},
\end{equation}
where $|q_1,q_2,... ,q_N\rangle$ denotes the computational basis state for $N$-qubit system, and $q$ denotes  $0$ or $1$ for each qubit. 
We parameterize the unitary transformation using elements of the Pauli group,
\begin{equation}
\hat{U}_N(\Theta) = {e}^{ -i \sum_g\theta_{g} \hat{P}_g },
\label{EqUN}
\end{equation}
where $\hat{P}_g$ represents an element of the Pauli group and is defined as
\begin{equation}
\hat{P}_g = \hat{\sigma}_{j_1}\otimes \hat{\sigma}_{j_2}\otimes ... \otimes\hat{\sigma}_{j_N},
\end{equation}
with $\hat{\sigma}_{j}$ denotes the identity ($j=0$) and the three Pauli matrices $(j=x,y,z)$. The index $g$ spans all possible configurations of $\{j_1,j_2,...,j_N\}$, excluding the trivial case where all $\hat{\sigma}_j$ are  identity matrices. The set of variational parameters is given by $ \Theta=\{\theta_g | g = 1,2,..., 4^N-1\}$. 
This universal  parameterization ensures the expressivity for  representing the diagonalizing transformation matrix. 

To identify the parameters in the unitary transformation, we utilize the summation of the squared probabilities of the computational-basis states as the objective function, which is written as
\begin{equation}
\mathcal{D}(\Theta)= \sum_{q_1,q_2,...,q_N} p_{q_1,q_2,...,q_N}^2.
\label{OBD}
\end{equation}
Noting that $p_{q_1,q_2,...q_N}$ is provided by the diagonal element of the evolved density matrix $\rho'$. When $\rho'$ is fully diagonalized, the objective function $\mathcal{D}$ becomes equivalent to  the purity of the input state,
\begin{equation}
\mathcal{P} = {\rm Tr}\{ \rho^2\},
\end{equation}
which is invariant under unitary transformations.  

\begin{figure}[t]
\begin{centering}
\includegraphics[width=0.48\textwidth]{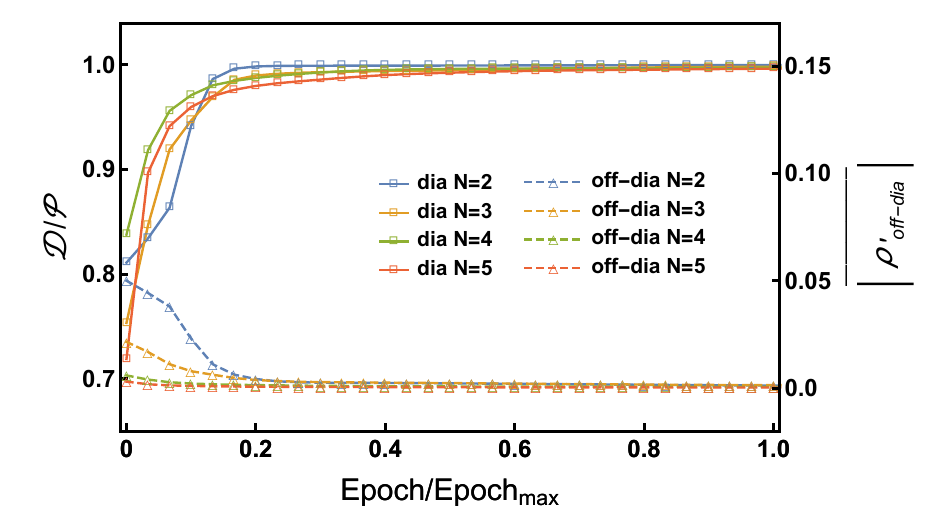}
\caption{ Training results for the variational quantum circuit with the objective function $\mathcal{D}$ for different qubit numbers $N$. (a) Throughout the optimization process, the ratio between objective function $\mathcal{D}$ and purity $\mathcal{P}$ is plotted over the training epochs as solid lines with square markers. 
(b) During training, $\overline{|\rho'_{\rm off-dia}|}$ is represented by a dashed line with triangular markers.  }
\label{Fig2}
\end{centering}
\end{figure}

With the goal of maximizing $\mathcal{D}$, the numerical training results for different qubit numbers, $N=2, 3, 4, 5$, are presented in Fig.~\ref{Fig2}.  For each value of $N$, we randomly generate an arbitrary input quantum state, which includes finite off-diagonal elements. In the  case of a pure input state, the spectral decomposition simplifies to a single dominant eigenvalue, allowing the state to be easily transformed into a product of single-qubit zero states \cite{Yao_2024}. However, for a general mixed quantum state, the spectral decomposition  becomes more complex due to the presence of multiple eigenvalues and the nontrivial distribution of diagonal elements. 
As shown in Fig.~\ref{Fig2}, the evolution of $\mathcal{D}$ during training is plotted by the solid line with square markers. As the training progresses over a sufficient number of epochs, the objective function $\mathcal{D}$ converges to its maximum, which corresponds to the expected purity value. 
During the training process, we also calculate the average amplitude of the off-diagonal elements of $\rho'$ as follows: 
\begin{equation}
\overline{|\rho'_{\rm off-dia}|} = \frac{1}{2^{N}(2^N-1)}\sum_{i\neq j }|\rho^\prime_{ij}|,
\label{Eqrhooff}
\end{equation}
which are plotted by the dashed lines with triangular markers. As illustrated in Fig.~\ref{Fig2}, at the end of each training, when $\mathcal{D} =\mathcal{P}$, all the off-diagonal elements vanish, indicating that the final state $\rho'$ obtains a diagonal form. 
Note that neither the purity nor $\overline{|\rho'_{\rm off-dia}|}$ are directly accessed during the optimization process, nor are they required for the variation of the quantum circuit. Experimental evaluation of purity and $\overline{|\rho'_{\rm off-dia}|}$ would typically involve quantum state tomography \cite{PhysRevLett.90.193601,PhysRevA.64.052312, MauroDAriano2003}. In our approach, however, only the objective function, provided by the computational-basis probabilities,  is required for optimization, which can be estimated statistically through repeated measurement shots.

As illustrated by the numerical results in Fig.~\ref{Fig2}, by maximizing $\mathcal{D}$ until it converges to the purity, we can transform the input quantum state into a diagonal form. This process reveals the eigenvalues and eigenvectors in quantum form: the probabilities of each computational basis state correspond to the eigenvalues of the input state, while the identified variational unitary provides the eigenvectors~\cite{VQdia}. In the following, we will analytically prove that $\mathcal{D}$ is upper-bounded by the purity, and when $\mathcal{D}=\mathcal{P}$, the associated quantum density matrix assumes a diagonal form with vanishing off-diagonal elements. 
Considering a general input quantum state $\rho$, which satisfies the Hermitian property $\rho = \rho^\dagger$,  the purity can be calculated as 
\begin{equation}
\mathcal{P} = \sum_{ij} \rho_{ij} \rho_{ji} =  \sum_{ij} \rho'_{ij} \rho'_{ji} = \sum_{i} \rho_{ii}^{\prime2} +\sum_{i\neq j} \rho'_{ij}\rho_{ij}^{\prime*},
 \label{EqP}
\end{equation}
where the second term is non-negative. This leads to the following inequality: 
\begin{equation}
 \mathcal{P} \geqslant  \sum_{i} \rho_{ii}^{\prime2} =\mathcal{D}.
 \label{EqPD}
\end{equation}
Here the equality holds if and only if  all off-diagonal elements vanish, ie.,  $\rho'_{i\neq j} =0$. 

In principle, the proposed variational quantum diagonalization algorithm can be implemented experimentally on a quantum platform. 
However, estimating the probabilities across the entire global computational basis results in a measurement complexity that scales exponentially with the number of qubits.
For an $N$-qubit quantum state, the total number of computational basis states is $2^N$. To achieve low statistical error in estimating the probabilities for all $2^N$ possibilities, the number of measurement shots required must be much larger than $2^N$. As a result, obtaining accurate estimates for each computational-basis probability becomes impractical as the system size $N$ increases.
To address this issue, we propose a second objective function based on local probabilities.

\subsection{local probabilities}
\subsubsection{Single-qubit case}
\begin{figure}[t]
\begin{centering}
\includegraphics[width=0.48\textwidth]{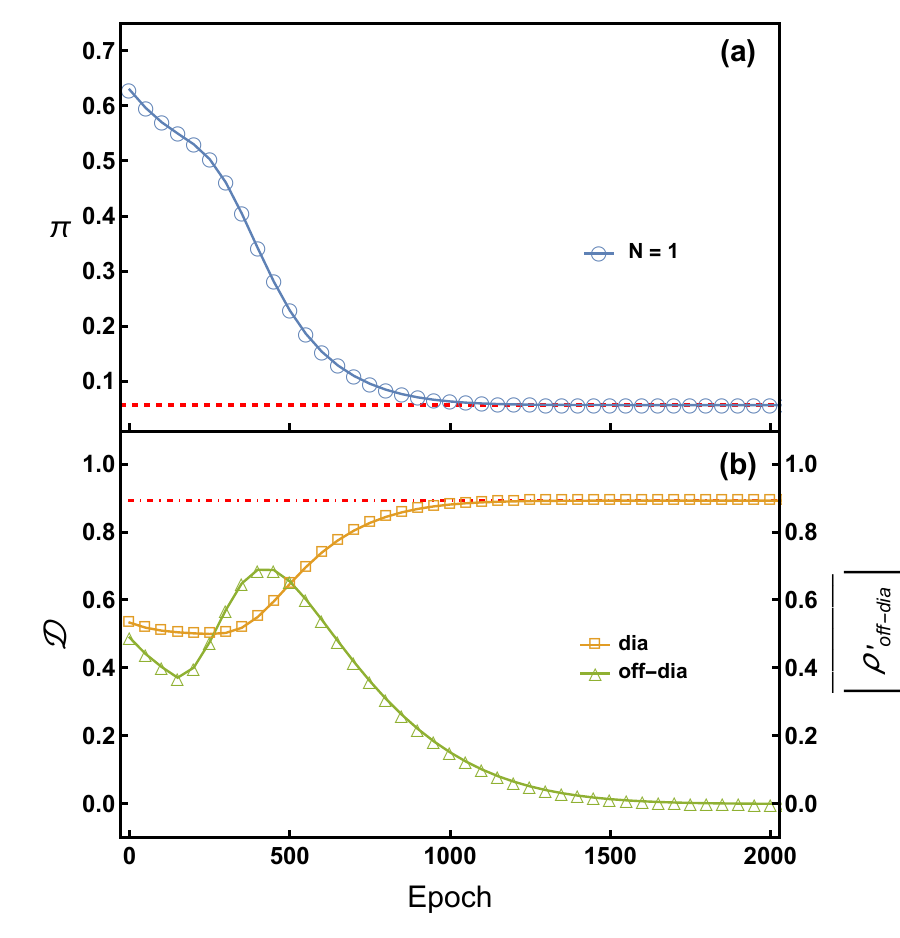}
\caption{Quantum diagonalization for an arbitrary $N=1$ quantum state using the single-qubit zero-state probability as the objective function. 
(a) During training, the zero-state probability $\pi$ is shown as a  blue line with circular markers. The red dashed line represents the smaller eigenvalue of the input density matrix $\rho$. 
(b) During training, $\mathcal{D}$ and $\overline{|\rho'_{\rm off-dia}|}$ are plotted as  lines with square and triangular markers, respectively.  The dot-dashed line indicates the purity $\mathcal{P}$ of the input quantum state.    }
\label{Fig3}
\end{centering}
\end{figure}
As a warm-up example, we consider the single-qubit case to illustrate the diagonalization process based on the zero-state probability. An arbitrary single-qubit quantum state can be represented by a density matrix $\rho$ in the form of 
\begin{equation}
\rho = 
\begin{bmatrix}
   \rho_{11} & \rho_{12} \\
\rho_{21} & \rho_{22}.
\end{bmatrix}.
\end{equation}
Here, the property of Hermitian and normalization condition imply that $\rho_{12} = \rho_{21}^*$ and $ \rho_{11} + \rho_{22} = 1$.
A general unitary transformation for a single qubit can be parametrized as:  
\begin{equation}
\hat{U}_{N=1}(\phi,\theta,\omega) = \begin{bmatrix}
\cos\frac{\theta}{2}e^{-i\frac{\phi+\omega}{2}}  &  -\sin\frac{\theta}{2}e^{i\frac{\phi-\omega}{2}}  \\
\sin\frac{\theta}{2}  e^{-i\frac{\phi-\omega}{2}} & \cos\frac{\theta}{2}e^{i\frac{\phi+\omega}{2}}  
\end{bmatrix},
\label{EqU1}
\end{equation}
where $\phi, \theta, \omega$ are the parameters that define the transformation. 
For optimization purposes, the objective function adopted here is the zero-state probability, defined as:
\begin{equation}
\pi = {\rm Tr}[\rho' |0\rangle\langle0|],
\label{EqpiN1}
\end{equation}
where $\rho'$ is the evolved quantum state given by $\rho'=\hat{U}\rho\hat{U}^\dagger$. 
The probability $\pi$ can be statistically estimated from outcomes of multiple measurement shots.  

The training process is numerically simulated by optimizing the parameters $\{\theta, \omega,\phi\}$ in Eq.~\eqref{EqU1}, with the goal of  minimizing the objective function $\pi$ of Eq.~\eqref{EqpiN1}. 
As shown in Fig.~\ref{Fig3}~(a), the zero-state probability $\pi$ decreases steadily throughout the training process, eventually converging to a minimum value.  At the end of the process,  the minimum value $\pi_{min}$ corresponds to the smaller eigenvalue of the input density matrix, as indicated by the red dashed line.  To monitor the optimization process, we evaluate both $\mathcal{D}$ in Eq.~\eqref{OBD} and $\overline{|\rho'_{\rm off-dia}|}$ in Eq.~\eqref{Eqrhooff}, as shown in Fig.~\ref{Fig3}~(b). After optimization, $\mathcal{D}$ saturates to the value of purity (orange line with squares), and all the off-diagonal elements of $\rho'$ vanish (green line with triangles). This confirms that the evolved state $\rho'$ attains a fully diagonal form by the end of the training process.

For the $N=1$ case, the zero-state probability $\pi$ is found to be bounded  by  the two eigenvalues of the input quantum state.  When the input quantum state is pure,  $\pi$ spans its maximum range, varying from $0$ to $1$. Minimization of $\pi$ will transform the input quantum state into state of $|1\rangle\langle 1|$. 
In contrast, for a mixed quantum state,  the presence of mixture reduces the range of $\pi$. 
To analytically explore the relationship between the range of $\pi$ and the evolution of the quantum state, the evolved quantum state $\rho'$ can be explicitly derived, allowing $\pi$ to be expressed as (see more details in method section):  
\begin{equation}
\begin{aligned}
\pi(\theta,\phi)=  &\frac{1}{2}+\frac{\rho_{11}-\rho_{22}}{2}\cos\theta \\
&- \frac{\rho_{12}+\rho_{21}}{2}\sin\theta\cos\phi\\
&-\frac{\rho_{12}-\rho_{21}}{2i} \sin\theta\sin\phi.
\end{aligned}
\end{equation}
Taking the derivatives of $\pi(\theta,\phi)$ with respect to $\theta$ and $\phi$, the local extremal values are given by:
\begin{equation}
\pi^{\pm} = \frac{1}{2}\pm \sqrt{ \frac{(\rho_{11}-\rho_{22})^2}{4}+\rho_{12}\rho_{21}},
\end{equation}
where $\pm$ corresponds to the local maximal and minimal value, respectively.   
It can be found that the two extreme values of $\pi$ correspond to the two eigenvalues of the input density matrix $\rho$. This result indicates that both maximizing and minimizing the zero-state probability can realize a diagonalization of the input quantum state. At these extremes, the off-diagonal elements can be calculated as $\rho_{12}^\prime=\rho_{21}^\prime=0$, ensuring that $\rho'$ attains a diagonal form. 
This indicates that the evolution of zero-state probability is constrained by two eigenvalues of the input quantum state. Meanwhile,  at the extremes, the evolved quantum state is diagonal. 

\subsubsection{Multi-qubit case}  
\begin{figure}[t]
\begin{centering}
\includegraphics[width=0.48\textwidth]{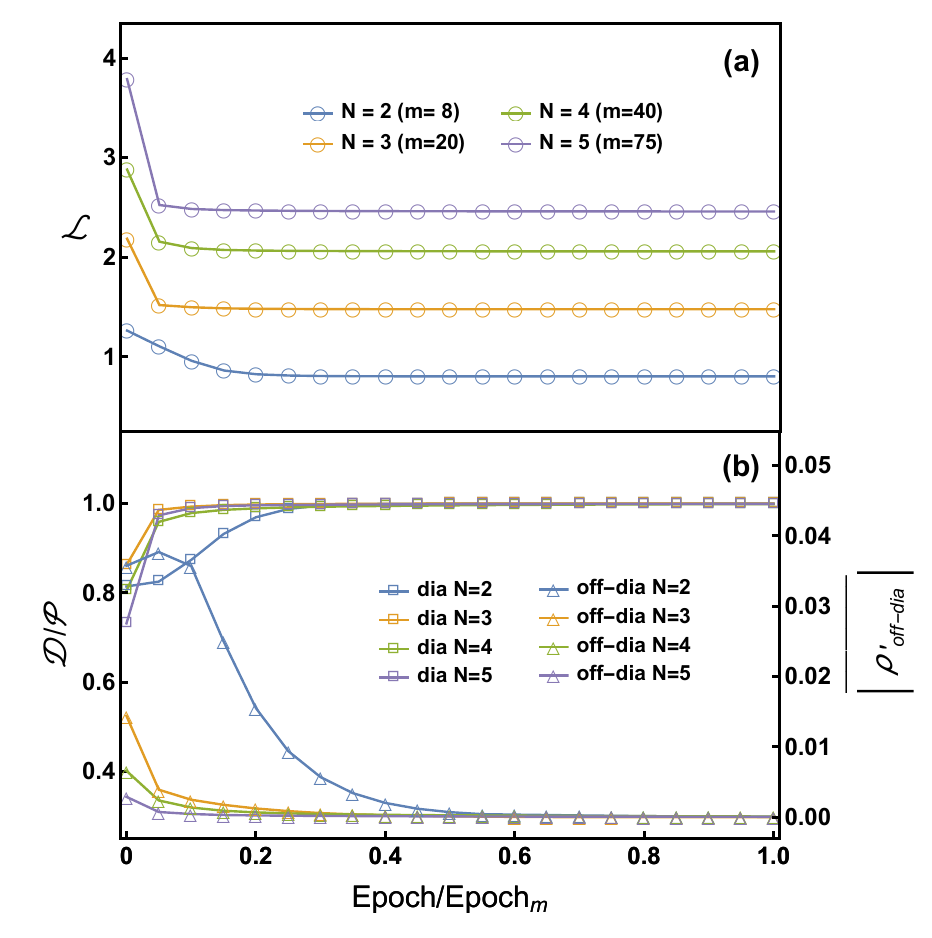}
\caption{Variation diagonalization results for various $N$-qubit quantum states with local computational-basis probabilities. (a) For each $N$, the objective function $\mathcal{L}$ is plotted as a line with circular markers. $m$ denotes the number of brick-wall building blocks for the corresponding quantum circuit. (b) For each optimization process, $\mathcal{D}$ is shown as a line with square markers, while the average magnitude of off-diagonal elements $\overline{|\rho^\prime_{\rm off-dia}|}$ is represented by  a line with triangular markers.   }
\label{Fig4}
\end{centering}
\end{figure}
Given the intriguing results observed in single-qubit systems, it is reasonable to hypothesize that there exists a connection between the single-qubit probabilities and diagonalization of an $N$-qubit quantum state. Motivated by this insight, we propose the following elegant and remarkably compact objective function, constructed solely from 
$N$ single-qubit probabilities at the zero state:
\begin{equation}
\mathcal{L} = \sum_{q=1}^{N}\sum_{n=1}^{N-q+1} \pi_q^n,
\end{equation}
where $\pi_q$ denotes the probability of the zero state for the $q$-th qubit, and $n$ represents the allowed power index associated with that single-qubit probability. Within the double summation, the maximal polynomial degree is $n=N$, which arises only when ${q=1}$; conversely, for $q=N$, the summation includes only the linear term $n=1$.  

This structure naturally reflects the hierarchical contribution of each qubit to the overall objective function.
Notably, when the system size is $N=1$, the objective function $\mathcal{L}$ reduces to a single probability term $\pi$, consistent with Eq.~\eqref{EqpiN1} for the single-qubit case. 
It is also important to emphasize that permuting any two qubits in the quantum state leaves the optimization process invariant. Consequently, the specific form of $\mathcal{L}$ is not unique for the purpose of optimization, and equivalent formulations can be constructed through qubit reordering.
This formulation highlights the expressive power of local observables in capturing essential global quantum structure, offering both conceptual clarity and computational efficiency. 
What makes this result particularly impressive is that such a compact and structured objective function, constructed purely from local single-qubit statistics, is nonetheless capable of driving a fundamentally global task for the diagonalization of an $N$-qubit quantum state.

To ensure the capability of the quantum circuit for diagonalization, a universal parameterization strategy can be adopted as given in Eq.~\eqref{EqUN}. However, this strategy involves multi-qubit interactions represented by the  product of non-trivial Pauli matrices. For experimental implementation, various sufficiently expressive quantum circuit architectures can also be used \cite{PhysRevResearch.3.L032057,QCarchitecturesearch}, provided they incorporate the desired diagonalizing unitary transformation.
Here, we construct a hardware-efficient quantum circuit using a brick-wall architecture, denoted as $\hat{\mathcal{B}}(\Theta_j)$ (see more details in method section). The variational quantum circuit is constructed with $m$ brick-wall building blocks, where the overall unitary transformation is given by $\hat{U}(\Theta)=\Pi_{j=1}^m \hat{\mathcal{B}}(\Theta_j)$. The number of building blocks $m$ required for each $N$ is determined dynamically during the  training process. The depth of the quantum circuit is extended until the optimization of the objective function converges, at which point the minimal value of $\mathcal{L}$ is reached in terms of $m$, as shown in Fig.~\ref{Fig4}~(a).
As training progresses, as shown in Fig.~\ref{Fig4}~(b), the quantity $\overline{|\rho'_{\text{off-dia}}|}$ becomes infinitesimally small, effectively vanishing all off-diagonal elements of the state. At this point, the diagonal form of the evolved state $\rho'$ is confirmed by $\mathcal{D}$ saturating to the purity $\mathcal{P}$.

Compared to the global-probability-based objective function $\mathcal{D}$  in Eq.~\eqref{OBD}, the local-probability objective function $\mathcal{L}$ offers a substantial reduction in measurement complexity.  Specifically, for each qubit, only two real parameters, the probabilities of obtaining $|0\rangle$ and $|1\rangle$, needed to be estimated, which corresponds to characterizing the reduced density matrix of a single qubit. In contrast, the global computational-basis probability distribution involves $2^N$ parameters, increasing exponentially with the number of qubits $N$.

Compared to global-probability objective function $\mathcal{D}$ of Eq.~\eqref{OBD}, the local-probability-based objective function $\mathcal{L}$ significantly reduces the required number of measurement shots.  For each qubit, only two degrees of freedom need to be statistically estimated. In contrast, the global computational basis probabilities involve $2^N$ degrees of freedom, growing exponentially with system size. 
As a result, the statistical resources required to evaluate $\mathcal{L}$ scale only linearly with system size, in stark contrast to the exponential scaling of $\mathcal{D}$. This renders the optimization of $\mathcal{L}$ far more feasible for large-scale quantum systems, both in simulation and on near-term quantum hardware \cite{doi:10.1126/science.abq5769}.

\section{Discussion}

In this report, we propose a variational quantum diagonalization algorithm based on computational-basis probabilities. Exploiting the connection between measurement probabilities and quantum state diagonalization, we introduce two objective functions to guide the optimization process.
The first objective function, $\mathcal{D}$, uses global computational-basis probabilities. By maximizing the sum of their squared amplitudes, this function drives the state toward a diagonal form. We analytically show that its optimal value corresponds to the purity of the input quantum state, offering a clear theoretical interpretation
However, estimating all  $2^N$ basis probabilities requires exponential measurement resources, making $\mathcal{D}$ impractical for large systems. To overcome this, we propose a second objective function, $\mathcal{L}$, based on local single-qubit probabilities. Remarkably, $\mathcal{L}$ captures essential features of the diagonalization task using only $N$ local observables. Despite its simplicity,  $\mathcal{L}$ effectively captures key aspects of the diagonalization task. It reveals a striking local-to-global correspondence: although constructed solely from local statistics, it guides the full quantum state toward diagonalization. The measurement complexity scales linearly with system size, offering a substantial advantage over the exponential cost of global approaches. This makes  especially attractive for near-term quantum devices \cite{doi:10.1126/science.abq5769}, where sampling efficiency is crucial.

While  performs well empirically, a rigorous theoretical understanding of $\mathcal{L}$ in multi-qubit systems remains open. In particular, how local information encodes the global diagonal structure deserves further exploration.
Nonetheless, the proposed variational framework combines efficiency, scalability, and physical interpretability. It offers a viable and experimentally feasible path for quantum state processing, with the potential to contribute to quantum advantage on noisy intermediate-scale quantum (NISQ) platforms \cite{preskill_quantum_2012, harrow_quantum_2017}.

\section{Methods}
\subsection{Analytic proof of $\lambda_-\leqslant\pi\leqslant \lambda_+$ for the single-qubit case}
For single qubit case, the evolved state can be explicitly derived according to $\rho' = \hat{U}\rho\hat{U}^\dagger$ with $\hat{U}_{N=1}$ parametrized by 
\begin{equation}
\hat{U}_{N=1}(\phi,\theta,\omega) = \begin{bmatrix}
\cos\frac{\theta}{2}e^{-i\frac{\phi+\omega}{2}}  &  -\sin\frac{\theta}{2}e^{i\frac{\phi-\omega}{2}}  \\
\sin\frac{\theta}{2}  e^{-i\frac{\phi-\omega}{2}} & \cos\frac{\theta}{2}e^{i\frac{\phi+\omega}{2}}  
\end{bmatrix}.
\end{equation}
The zero-state probability is given by the diagonal elements of $\rho'$, which can be explicitly derived as 
\begin{equation}
\begin{aligned}
\rho^\prime_{11}(\theta,\phi) = &\frac{1}{2}+\frac{\rho_{11}-\rho_{22}}{2}\cos\theta \\
&- \frac{\rho_{12}+\rho_{21}}{2}\sin\theta\cos\phi -\frac{\rho_{12}-\rho_{21}}{2i} \sin\theta\sin\phi,
\end{aligned}
\label{EqA1}
\end{equation}
and the off-diagonal element is given by 
\begin{equation}
\begin{aligned}
\rho^\prime_{12}(\theta,\phi,\omega) = &\frac{ \rho_{11}-\rho_{22} }{2 e^{i\omega}} \sin\theta + \rho_{12}\frac{1+\cos\theta}{2 e^{i\omega}} e^{- i \phi} \\
&- \rho_{21}\frac{1-\cos\theta}{2 e^{i\omega}}e^{ i \phi} .
\end{aligned}
\end{equation}
The local extrema correspond to points where the derivatives are zero.
Setting the first derivative of $\rho'_{11}$ with respect to $\phi$ equals to zero yields, 
\begin{equation}
R \sin\theta \sin\phi = I \sin\theta\cos\phi,
\label{EqDphi}
\end{equation}
where denoting $R=( \rho_{12}+\rho_{21})/2$ for the real part and $I = ( \rho_{12}-\rho_{21})/2i$ for the imaginary part. Here we assume both the real and imaginary part are finite for the off-diagonal element of the input quantum state. Under this assumption, the solution for $\phi$ can be casted as 
\begin{equation}
\begin{aligned}
\cos\phi^* = \pm \frac{R}{\sqrt{R^2+I^2}}, \\
\sin\phi^* = \pm \frac{I}{\sqrt{R^2+I^2}}. 
\end{aligned}
\end{equation}
Then inserting the above solution for $\phi = \phi^*$  into Eq.~\eqref{EqA1}, we arrive 
\begin{equation}
\begin{aligned}
\rho^\prime_{11}(\theta,\phi^*) = &\frac{1}{2}+\frac{\rho_{11}-\rho_{22}}{2}\cos\theta  \mp\sqrt{R^2+I^2}\sin\theta,
\end{aligned}
\end{equation}
and the off-diagonal element
\begin{equation}
\begin{aligned}
\rho^\prime_{12}(\theta,\phi^*,\omega)e^{i\omega} = &\frac{ \rho_{11}-\rho_{22} }{2} \sin\theta -iR\sin\phi+iI\cos\phi  \\
& +R\cos\theta\cos\phi+I\cos\theta\sin\phi \\
=&\frac{ \rho_{11}-\rho_{22} }{2} \sin\theta \pm \sqrt{R^2+I^2}\cos\theta. 
\end{aligned}
\end{equation}
Then derivative of $\rho_{11}'$ in terms of $\theta$, $\partial_\theta \rho'_{11}=0$,  leads to 
\begin{equation}
\begin{aligned}
\mp \sqrt{4R^2+4I^2} \cos\theta = (\rho_{11}-\rho_{22})\sin\theta,
\end{aligned}
\end{equation}
which results in
\begin{equation}
\begin{aligned}
\cos\theta^* = \pm \frac{\rho_{11}-\rho_{22}}{\sqrt{(\rho_{11}-\rho_{22})^2+4R^2+4I^2}}, \\
\sin\theta^* = \pm \frac{\mp\sqrt{4R^2+4I^2}}{\sqrt{(\rho_{11}-\rho_{22})^2+4R^2+4I^2}}. 
\end{aligned}
\end{equation}
Solutions for the $\phi^*$ and $\theta^*$ provides the result of $\rho'$ at the extreme points with 
\begin{equation}
\begin{aligned}
\rho^\prime_{11}(\theta^*,\phi^*) &= \frac{1}{2}\pm \frac{1}{2} \sqrt{ (\rho_{11}-\rho_{22})^2+4R^2+4I^2} \\
&= \frac{1}{2}\pm \frac{1}{2} \sqrt{ (\rho_{11}-\rho_{22})^2+4\rho_{12}\rho_{21}}, \\
\rho^\prime_{12}(\theta^*,\phi^*,&\omega ) = 0 . 
\end{aligned}
\label{EqExt}
\end{equation}
Here $\rho'_{11}(\theta^*, \phi^*)\equiv \pi^{\pm}$ provide the extremes for zero-state probability, with $\pi^-\leqslant\pi\equiv\rho'_{11}(\theta, \phi)\leqslant \pi^+$. Meanwhile, the eigenvalues of $\rho$, through solving the characteristic equation ${\rm det}(\rho-\lambda I)=0$, can be derived as  
\begin{equation}
 \lambda^{\pm} = \frac{1}{2}\pm \sqrt{ \frac{(\rho_{11}-\rho_{22})^2}{4}+\rho_{12}\rho_{21}},
\end{equation}
which equal to the extremes $\pi^{\pm}$ calculated by Eq.~\eqref{EqExt}. This implies that the zero-state probability is constrained between the two eigenvalues of the input quantum state, 
\begin{equation}
\lambda_-\leqslant\pi\leqslant \lambda_+ ,
\end{equation}
with equality holding when the corresponding evolved state is diagonal.

\subsection{Optimization with hardware-efficient quantum circuit}
\vspace{5mm}
\begin{figure}[t]
\begin{centering}
\includegraphics[width=0.4\textwidth]{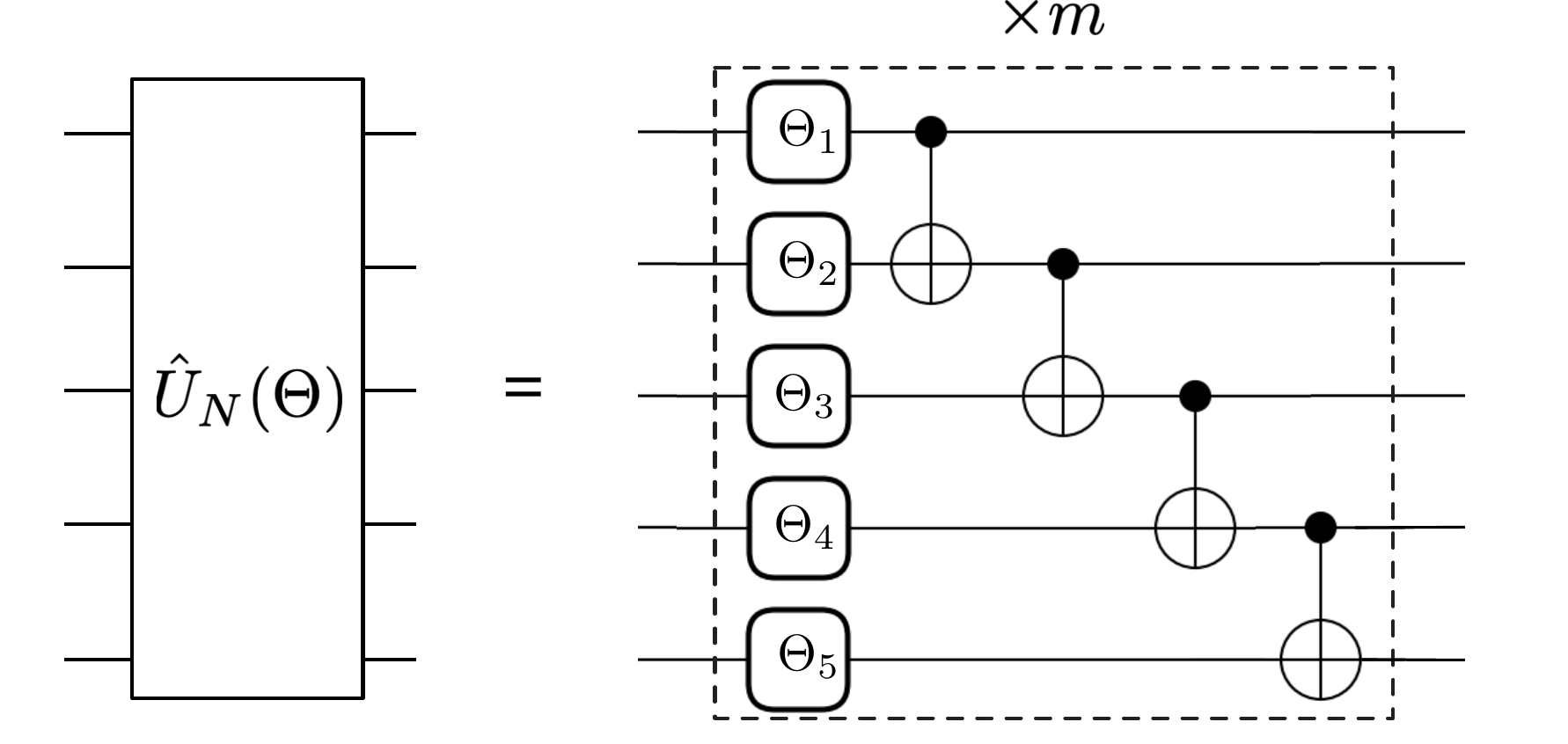}
\caption{ For an $N=5$ qubit system, the quantum circuit is constructed using $m$ building blocks. As illustrated within the dashed box, the building block $\hat{\mathcal{B}}(\Theta)$ follows a brick-wall architecture.}
\label{FigAppBW}
\end{centering}
\end{figure}
\begin{figure}[h!!]
\begin{centering}
\includegraphics[width=0.5\textwidth]{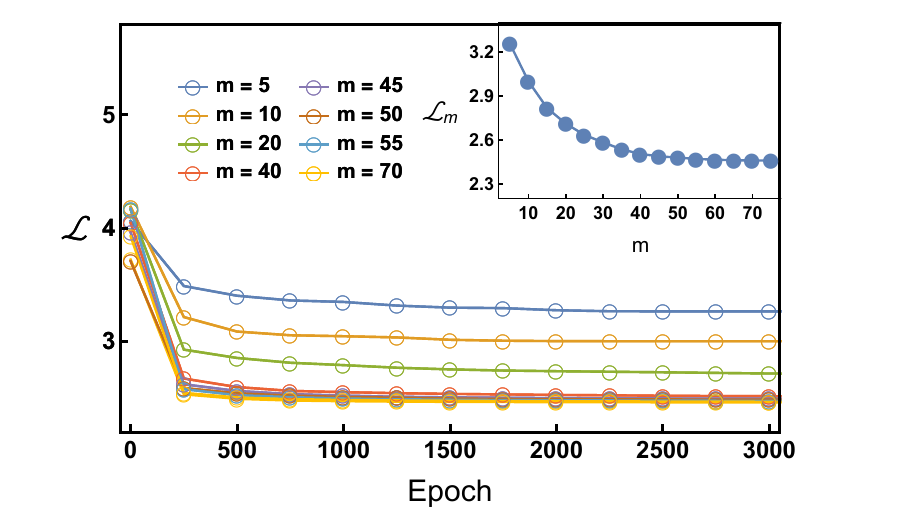}
\caption{ Dynamic training process for a $5$-qubit quantum state using the brick-wall building block.
For each $m$, the objective function $\mathcal{L}$ is plotted as a line with circular markers. The inset shows the minimal objective function $\mathcal{L}_m$ obtained for each $m$. }
\label{FigApp2Train}
\end{centering}
\end{figure}
\begin{figure}[t!!]
\begin{centering}
\includegraphics[width=0.5\textwidth]{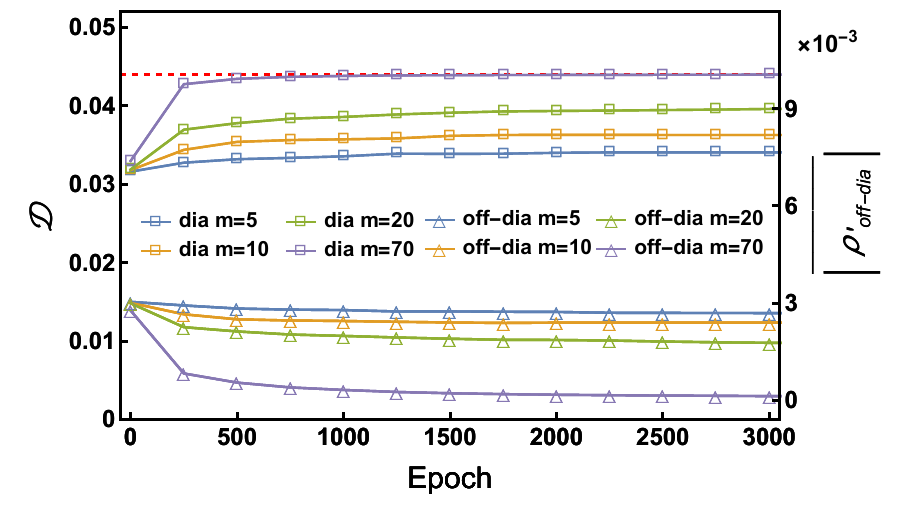}
\caption{ For each $m$, the diagonal quantities $\mathcal{D}$ and off-diagonal quantities $\overline{|\rho^\prime_{\rm off-dia}|}$ are represented are shown as  lines with triangular and square markers respectively. The red dashed line indicates the purity of the quantum state with $5$ qubits.  }
\label{FigApp3Train}
\end{centering}
\end{figure}
\vspace{5mm}
The numerical simulations are performed using PyTorch \cite{paszke_pytorch_2019} and PennyLane \cite{bergholm_pennylane_2022}, with optimization carried out using the Adam optimizer. 
A universal parameterization strategy, as described in Eq.~\eqref{EqUN}, ensures the capability for quantum state diagonalization. However, this approach relies on multi-qubit gates, which are challenging for experimental implementation. To address this, we adopt a hardware-efficient quantum circuit based on a brick-wall architecture. The expressivity of the quantum circuit is ensured by increasing the circuit depth dynamically during training.
As illustrated in Fig.~\ref{FigAppBW}, the brick-wall building block, $\hat{\mathcal{B}}(\Theta)$,  begins with $N$ single-qubit gates, followed by $N-1$ CNOT gates. The single-qubit gates are defined by the expression in Eq.~\eqref{EqU1}, where each set of parameters $\Theta_q$ contains three elements: $\theta,\omega$ and $\phi$. 
This building block is then repeated $m$ times to construct a variational quantum circuit , expressed as  
\begin{equation}
\hat{U}_{N}(\Theta)=\Pi_{j=1}^m \hat{\mathcal{B}}(\Theta_j),
\end{equation}
as shown in Fig.~\ref{FigAppBW}. 

Taking $N=5$ as an illustration,  we will explain how to determine the required number of $m$ or the depth of the quantum circuit for the diagonalization purpose.  As shown in Fig.~\ref{FigApp2Train}, we begin with a shallow quantum circuit, choosing $m=5$ as an example.  As training progresses, shown by the blue line with open circular markers, the objective function  $\mathcal{L}$ approaches its minimum value $\mathcal{L}_m$ .  
As shown by the blue lines in Fig.~\ref{FigApp3Train}, in the end of training for $m=5$, the deviation of $\mathcal{D}$ from the purity (red dashed line) exists due to the presence of finite off-diagonal elements. This indicates that the corresponding quantum circuit with $m=5$ is too shallow to fully express the diagonal matrix. To address this, additional building blocks are incorporated into the circuit. 
The depth of the quantum circuit is increased by appending additional building blocks until $\mathcal{L}_m$  converges, as shown in the inset of Fig.~\ref{FigApp2Train}.  As indicated by the purple lines in Fig.~\ref{FigApp3Train}, when $m=70$,  the off-diagonal elements of the evolved state become infinitesimally small. Meanwhile, $\mathcal{D}$ saturates to the purity $\mathcal{P}$, confirming that the evolved state is diagonal.

\section{Acknowledgements}
We thank Hui Zhai, Yadong Wu, Zhigang Wu, Jiang Zhang, and Youpeng Zhong for helpful discussions. 
This work is supported by the Science, Technology and Innovation Commission of Shenzhen, Municipality (KQTD20210811090049034), Guangdong Basic and Applied Basic Research Foundation (2022B1515120021).
 
\section{Data availability}
Data generated and analyzed during the current study are available from https://github.com/jy19Phy/VQSD.
\section{Code availability}
Code used for the current study is available from https://github.com/jy19Phy/VQSD.

\bibliographystyle{unsrt}  
\bibliography{Ref.bib}


\end{document}